\documentclass[%
 preprint,
 amsmath,amssymb,
 aps,
 pre,
 onecolumn,
 superscriptaddress,
]{revtex4-2}

\usepackage{graphicx}%
\usepackage{dcolumn}%
\usepackage{bm}%

\DeclareMathOperator{\sinc}{sinc}

\DeclareMathOperator{\arcsinh}{arsinh}

\usepackage{stackengine}
\usepackage{mathtools}

\begin{document}

\title{Non-reciprocal anti-aligning active mixtures: deriving the exact Boltzmann collision operator}
\author{Jakob Mihatsch}
\affiliation{Institut f{\"u}r Physik,
Universit{\"a}t Greifswald, Felix-Hausdorff-Str. 6, D-17489 Greifswald, Germany}
\affiliation{Institut f{\"u}r Physik, Otto-von-Guericke-Universit{\"a}t Magdeburg, Universit{\"a}tsplatz 2, D-39106 Magdeburg, Germany}
\author{Thomas Ihle}
\affiliation{Institut f{\"u}r Physik,
Universit{\"a}t Greifswald, Felix-Hausdorff-Str. 6, D-17489 Greifswald, Germany}

\begin{abstract}
    We consider the effect of non-reciprocity in a binary mixture of self-propelled particles with
    anti-aligning interactions, where a particle of type A reacts differently to a particle of type B than
    vice versa.  Starting from a well-known microscopic Langevin-model for the particles,
    setting up the corresponding exact N-particle Fokker-Planck equation and making Boltzmann's assumptions of low density and
    one-sided molecular chaos, the non-linear active Boltzmann equation
    with the exact collision operator is derived. In this derivation, the effect of phase-space
    compression and the build-up of pair-correlations during binary interactions is explicitly taken into account, leading to
    a theoretical description beyond mean-field.
    This extends previous results for
    reciprocal interactions, where it was found that orientational order can emerge in a system with
    purely anti-aligning interactions. Although the equations of motion are more complex than in the
    reciprocal system, the theory still leads to analytical expressions and predictions. Comparisons with
    agent-based simulations show excellent quantitative agreement of the dynamic and static behavior
    in the low density and/or small coupling limit.
\end{abstract}

\maketitle

\section{Introduction}
Self-propelled particles show many interesting effects due to them being intrinsically far from equilibrium \cite{menzel2015tuned}.
Prominent examples are long range orientational order in two dimensions \cite{toner2005hydrodynamics} or motility induced phase separation \cite{cates2015motility}.
Another consequence of their activity is that interactions between those particles must not necessarily be reciprocal \cite{dinelli2023non}.
For example, in the Vicsek model, the averaging of directions over the neighboring particles introduces a non-reciprocity.
A more obvious way to introduce non-reciprocity is by including two species in the model, with a non-reciprocal strength of the coupling between them \cite{kreienkamp2022clustering,saha2020scalar}.

Binary mixtures with non-reciprocal interactions were brought forward as an example of a new kind of phase transition \cite{fruchart2021non}.
Inside the ordered phase, time-dependent steady states emerge. This is observed when the species have competing goals, species A wants to align polarly with species B, while species B wants to anti-align with species A.
Another recent finding is that non-reciprocal interactions lead to an asymmetric clustering in mixtures of active Brownian particles \cite{kreienkamp2024nonreciprocal}.

Several authors have also studied binary mixtures of dry aligning active matter with reciprocal interaction.
Studies focus mostly on species differing by the nature and strength of the interaction \cite{menzel2012collective,chatterjee2023flocking} or the noise affecting the interaction \cite{ariel2015order}.
These publications have in common that polar aligning interactions act among the same species, but the interactions between different species could lead to anti-alignment or even perpendicular velocities.
It was found that one species in a flocking state can induce polar order in a second species that would not be in a flocking state if it were on its own \cite{menzel2012collective,ariel2015order}.

Recently it was observed that even a binary system with only anti-aligning interactions can show a transition to flocking \cite{kursten2023flocking}.
This transition is caused by a frustration of the system.
The anti-alignment among the same species competes with anti-alignment between the two species.
Like the Vicsek model, the homogeneous polar ordered state in the binary anti-aligning system is unstable to long wavelength perturbations.
Unlike the Vicsek model, these instabilities do not only occur close to the transition, but anywhere in the ordered phase.
In some circumstances, these spatial inhomogeneities lead to a partial de-mixing of the two species.

One of the fundamental postulates in physics is Newton's third law ``actio=reactio", which states that the mechanical
force $\vec{F}_{ij}$ particle $j$ exerts
on particle $i$ is opposite to the force particle $i$ exerts on $j$: $\vec{F}_{ji}=- \vec{F}_{ij}$.
This leads to the conservation of total momentum of an isolated system of particles interacting via such pairwise additive forces.
The action-reaction symmetry and thus the conservation of particle momentum can be broken if the interaction among the particles
is mediated by some non-equilibrium environment, whereas, of course the momentum of the complete system ``particles-plus-environment''
must be conserved. During the last decade, scientists became very interested in the effects of such non-reciprocal forces in
complex plasmas \cite{lin2018structure,lisin2020experimental,zampetaki2020buckling}, metamaterials \cite{brandenbourger2019non,wang2023non,fleury2014sound,miri2019exceptional,scheibner2020odd,coulais2017static}, networks of
neurons \cite{sompolinsky1986temporal,montbrio2018kuramoto,martorell2024dynamically}, social groups and active particle systems, see for example Refs. \cite{ivlev2015statistical,fruchart2021non}. 
Many active particle systems consist of overdamped particles with constant speed $v_0$, which
exert torques on each other. As a result, for example in two dimensions, their dynamics is often described
by a differential equation for their flying angle
$\theta_i$ as
\begin{equation}
\dot{\theta_i}(t)=\sum_{j=1}^N\,H_{ij} \;\;\;{\rm with}\ H_{ij}\equiv H(\theta_j-\theta_i,\vec{r}_j-\vec{r}_i)
\end{equation}
plus some external noise which we neglect for now.
This equation contains a sum of pairwise interaction terms $H_{ij}$ that describe how particle $j$ affects the angle of particle $i$.
In analogy to the case of regular particles with actual forces, reciprocal interactions fulfill the rule
$H_{ji}=- H_{ij}$, something which is realized in one of the most commonly used models for active particles -- the Kuramoto model --
\begin{equation}
H_{ij}=\Gamma a_{ij}\ {\rm sin}(\theta_j-\theta_i),
\end{equation}
where $\Gamma$ is a constant indicating the strength of coupling and $a_{ij}=1$ for $|\vec{r}_j-\vec{r}_i|\leq R$ and $a_{ij}=0$ otherwise.
In an isolated system of such particles, it is not the total momentum that is conserved but
the rather unintuitive quantity $ S=\sum_{i=1}^N\theta_i$, the sum of all flying angles.

\begin{figure}
    \includegraphics[width=.45\textwidth]{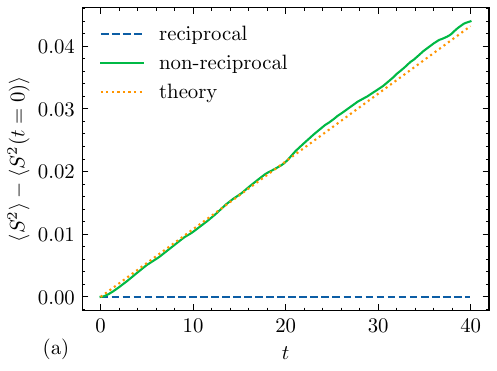}
    \includegraphics[width=.45\textwidth]{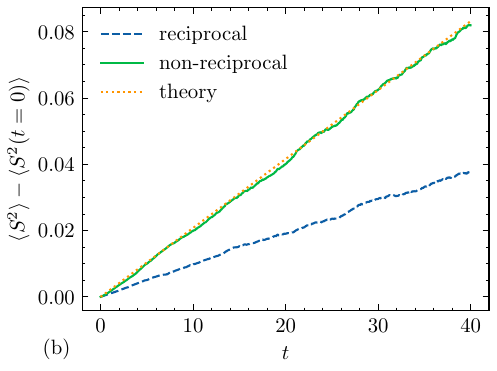}
    \caption{Time evolution of the variance of the sum of all directional angles $\langle S^2\rangle$. %
    For reciprocal interactions (blue dashed) and zero noise, $S$ is a conserved quantity. In the case of non-reciprocal interactions (green), it fluctuates strongly.
    Non-reciprocity is introduced by having two different species of particles (1 and 2), where the coupling constant $\Gamma_{\sigma_i,\sigma_j}$ between particle $i$ and $j$ now depends on the species $\sigma_i,\sigma_j\in{1,2}$ of the interacting particles.
    Simulation parameters are $Sc(1,1)=Sc(2,2)=0.02$, furthermore $Sc(1,2)=Sc(2,1) = 0.015$ in the reciprocal case, $Sc(1,2) = 0.015$, $Sc(2,1) = 0.005$ in the non-reciprocal case. 
    The coupling strength is defined as $Sc(\alpha,\beta)=-\frac{\Gamma_{\alpha,\beta}R}{v_0}$.  
    The particle number is $N_1=N_2=500$, the scaled density $M_1=M_2=0.03$ and $R=v_0=1$. The integration timestep is $\Delta t = 0.02$.
    The agent based simulations are compared to an analytical expression (orange) derived further below, see Eq. \eqref{eq:deltaSsqu}.
    Subfigure (a) shows the situation without external noise. In (b) there is some external noise $\sqrt{2}\nu=10^{-3}$. As $\nu\ll \min(1,Sc)$, this noise has no influence on a collision of two particles. In this situation the non-reciprocity causes an additional change in $S$ that is independent from the stochastic noise term.}
    \label{fig:sum}
\end{figure}

Thus, there seems to be some hidden symmetry whose physical implications are not clear.
In this paper, we would like to investigate what happens if this symmetry is strongly broken,
that is, if $|H_{ij}+H_{ji}|$ is nonzero and rather large. Fig. \ref{fig:sum} shows the variance angle sum $\langle S^2\rangle$ as a function of time
for a system of active particles with reciprocal and non-reciprocal Kuramoto alignment, making the difference between
the two types of systems quite visible.
In models without external noise, $S$ is strictly constant for reciprocal
versions of the model with $H_{ij}=-H_{ji}$. We expect that even with small external noise there
 will still be significant differences between the behavior of systems with reciprocal and non-reciprocal interactions.
We would like to set up a kinetic theory that delivers quantitative predictions
and is sensitive enough to capture the effects of this symmetry breaking.
Here, for now, we derive a kinetic equation that is valid for small external noise and focus our analysis of it on the most extreme case of zero external noise, leaving the detailed investigation
of nonzero noise for future work.
Breaking the hidden $S=const.$ symmetry might lead to rather subtle effects which is why we will develop and employ a Boltzmann theory
with exact collision operator instead of mean-field theory or other more ad hoc Boltzmann approaches \cite{bertin2009hydrodynamic,bertin2006boltzmann,peshkov2014boltzmann}. %
\footnotetext[5]{See for example Eq. S 58 in the supplemental of \cite{fruchart2021non}. It shows only the coupling of $f_{n-1}$ and $f_{n+1}$
to the modes $f_1$ and $f_{-1}$ which is a hallmark of the mean-field approximation.
Beyond mean field, infinitely more couplings occur, see Eq. \eqref{eq:nonrecmain:final}, in particular to the density mode $f_0$ }
Note that the kinetic and hydrodynamic equations of Fruchart et al. \cite{fruchart2021non,Note5} 
which are their basis of exploring the effects of non-reciprocity, are all based
on a simple mean-field approximation that is known to make unphysical predictions
for certain active particle systems \cite{ihle2023scattering,das2024flocking}.
In this paper, we aim to construct theories beyond mean field that can also faithfully handle the non-reciprocal versions of
those systems.

The theoretical description of self-propelled particles often relies on hydrodynamic equations that are either postulated on grounds of symmetry and contain phenomenological parameters \cite{toner1998flocks,toner2012reanalysis} %
or are derived from kinetic theories \cite{kursten2023flocking,baskaran2008hydrodynamics,bertin2006boltzmann,peshkov2014boltzmann}.
In many cases, the underlying kinetic description is achieved
via the mean-field (also called {\em Vlasov}- or simple molecular chaos-) assumption \cite{peruani2008mean,baskaran2008hydrodynamics}
which assumes particles as
statistically independent at {\em all times} and that their joint probability distributions factorizes.
For certain microscopic models of active particles, for example, the
``flocking by turning away''-model of Ref. \cite{das2024flocking}
or the noiseless anti-alignment models of Refs. \cite{kursten2023flocking,ihle2023scattering} %
this mean-field assumption fails dramatically and leads to unphysical predictions such the absence of a phase transition
or an infinite self-diffusion coefficient. Thus, it is useful to propose kinetic theories, as done in this paper, that do not
rely on the simple mean-field assumption.
A way to circumvent the usage of this assumption is to focus on models
where interactions occur only at discrete times, such as the standard Vicsek-model (VM) \cite{vicsek1995novel,vicsek2012collective},
and to ensure that particles have distanced themselves far enough from their collision partner in the
time interval following their encounter. Factorizing the joint probabilities of the particles at the discrete collision times
now effectively constitutes the far less drastic approximation of {\em one-sided molecular chaos}, where particles are only assumed
to be independant {\em before} an interaction but not immediatly afterwards. The Enskog-like kinetic theory for the standard Vicsek model \cite{ihle2011kinetic,ihle2016chapman} %
made use of this idea,
kept the discrete time evolution and was able to quantitatively describe the microphase separation and the density waves of the VM
in the limit of large mean free path \cite{ihle2013invasion}. %

The Boltzmann description of Bertin, Droz and Gregoire (BDG) \cite{bertin2006boltzmann,bertin2009hydrodynamic,peshkov2014boltzmann} %
for dilute systems also utilized
the idea of permitting interactions only at particular points in time,
allowing the implicit incorporation of one-sided molecular chaos in a simple fashion.
This kinetic approach, however, does neither follow from the collision rules of the standard or original VM
nor from simple microscopic Langevin-models with alignment \footnote{A possible corresponding microscopic model would have to contain a Maxwellian demon, who abruptly
forbids interactions once two particles have aligned in the Vicsek-sense and are still in interaction range 
but allows them to interact again once the particles have
been separated for a while and meet once again. Such a model would be unnecessarily complicated and hard to motivate physically.}, %
and thus agrees only qualitatively
but not quantitatively with popular microscopic and computational models for polar and nematic active matter.
In that sense, its derivation from Vicsek-like microscopic equations appears to be heuristic.

Apart from not promising quantitative results, the BDG approach is based on the instantaneous alignment
concept of the VM without an explicit parameter for alignment strength.
Therefore, it cannot capture the effect of different coupling strengths, which is important to describe
non-reciprocal interactions. Only very recently a Boltzmann equation for active systems was
proposed that is derived via the BBKGY hierarchy for a specific microscopic Langevin-model by using
the assumption of one-sided molecular chaos \cite{ihle2023scattering,ihle2023asymptotically} %
in a way that explicitly takes the build-up of particle correlations
during interactions into account.
This approach leads to the exact Boltzmann collision operator for the underlying microscopic dynamics.
Hence, it is asymptotically accurate in the low density limit and has already been applied in Ref. \cite{kursten2023flocking} %
to binary mixtures with
reciprocal interactions. There it was found that this active Boltzmann theory quantitatively
captures the dynamics and the stationary state of the system much more accurately than
the mean field description. It also yielded qualitatively new results, like a dependence of the
transition to polar order on the speed of the particles. More dramatically, the new theory yields the correct finite coefficient of self-diffusion that the mean-field approach falsely
predicts to be infinite in the limit of zero angular noise \cite{ihle2023asymptotically}. %
Very recently,  another approach,  a Landau kinetic theory, valid for small coupling strengths,  which also utilizes the
one-sided molecular chaos  assumption in its derivation from
microscopic  models with nematic alignment,  was  proposed
in Ref. \cite{boltz2024kinetic}. %

In this publication, we will apply this active Boltzmann approach to a binary mixture with non-reciprocal anti-aligning interactions, extending the results described in Ref. \cite{kursten2023flocking} where only reciprocal interactions were investigated.

\section{Active Boltzmann equation}
We use the following active Brownian model for point-like particles with Kuramoto-alignment (found e.g. in Refs. \cite{peruani2008mean,menzel2012collective,kursten2023flocking,fruchart2021non})
\begin{subequations}
\begin{eqnarray}
    &\dot{\vec{r}}_j = v_0\hat{n}(\theta_j) \\
    &\dot{\theta}_j = \sum_{k,|\vec{r}_j-\vec{r}_k|<R} \Gamma_{\sigma_j, \sigma_k} \sin(\theta_k-\theta_j) + \sqrt{2}\,\nu \eta_j\\
    &\langle \eta_j(t)\rangle=0,\ \langle \eta_j(t)\eta_k(t')\rangle=\delta(t-t')\delta_{jk}.
\end{eqnarray}
\label{eq:DAAM:rulespoly}
\end{subequations}
It describes the evolution of the particles' position $\vec{r}_j$ in 2D-space and of the angle $\theta_j$ indicating their direction of motion.
The indices $\sigma_1,...,\sigma_N$ identify the species of the particles, which in this case is either type 1 or type 2.
Here, the coupling constant depends on the type of particle and is given by the constants $\Gamma_{\sigma,\tilde{\sigma}},\ \sigma,\tilde{\sigma}\in\left\{1, 2\right\}$.
The species only differ by the interaction constant, but all move with the same velocity, have the same interaction range and are affected by a noise of the same strength. 
The species can also have a different particle number $N(\sigma)$.
There are three obvious timescales involved: The typical collision timescale $R/v_0$, the timescale $1/|\Gamma|$ on which the alignment occurs and the timescale of the angular diffusion due to the noise $1/\nu^2$.
There is also an emergent time scale, the decay time of the velocity auto-correlations in a noise-free
system, that scales as $v_0/(R^3 \Gamma^2 \rho_N)$ with number density $\rho_N$ as shown in Ref. \cite{ihle2023asymptotically}, see
also Eq. (32) \footnote{This time plays the role of an effective mean free time between randomizing collisions and
is also proportional to the inverse decay rate of higher angular modes.}.
In a system with polar aligning interactions, the noise and the alignment are competing.
We therefore keep these parameters and eliminate $v_0$ and $R$ by rescaling for times $t$ and lengths $x$.
\begin{equation}
    \begin{aligned}
        t&\rightarrow \bar{t}\cdot \frac{R}{v_0}\\*
        x&\rightarrow \bar{x}\cdot R,
        \label{eq:kinetic:units}
    \end{aligned}
\end{equation}
effectively setting $v_0=R=1$. To simplify notations we will again drop the line over the dimensionless quantities in the following. Eq.~\eqref{eq:DAAM:rulespoly} reads in dimensionless form
\begin{subequations}
    \begin{eqnarray}
        &\dot{\vec{r}}_j = \hat{n}(\theta_j) \\*
        &\dot{\theta}_j = - \sum_{k,|\vec{r}_j-\vec{r}_k|<1} Sc(\sigma_j, \sigma_k) \sin(\theta_k-\theta_j) + \sqrt{2}\,\bar{\nu}\bar{\eta}_j\\*
        &\langle \bar{\eta}_j(t)\rangle=0,\ \langle \bar{\eta}_j(t)\bar{\eta}_k(t')\rangle=\delta(t-t')\delta_{jk}
    \end{eqnarray}
    \label{eq:kinetic:dimensionlessrules}
\end{subequations}
with the dimensionless coupling strength $Sc(\sigma_j, \sigma_k)=-\frac{R}{v_0}\Gamma_{\sigma_j, \sigma_k}$, the dimensionless noise strength $\bar{\nu}^2=\frac{\nu^2R}{v_0}$ and the rescaled noise $\bar{\eta}=\eta\sqrt{\frac{R}{v_0}}$.
As before we will not write the line over the symbols in the following chapters to simplify notations.
Note that $Sc$ was defined in such a way that it is positive for anti-aligning interactions, the case predominantly treated in this paper.
The aligning case can however still be recovered by allowing $Sc<0$.
The particles are placed in a square box with length $L$ and periodic boundary conditions.
The dimensionless number density is defined by $\rho=\frac{NR^2}{L^2}$, and will often be expressed by the expected number of neighbors $M=\rho\pi$.

We now assume that in general
\begin{equation}
        Sc(\sigma,\bar{\sigma})\neq Sc(\bar{\sigma},\sigma)
\end{equation}
and define
\begin{eqnarray}
    Sc^+(\sigma,\bar{\sigma})= Sc(\bar{\sigma},\sigma)+Sc(\sigma,\bar{\sigma})
\end{eqnarray}
\begin{eqnarray}
    Sc^-(\sigma,\bar{\sigma})=Sc(\bar{\sigma},\sigma)-Sc(\sigma,\bar{\sigma}).
\end{eqnarray}
$Sc^+$ is a measure of the strength of the interaction and still required to be positive, so that we only deal with overall anti-aligning interactions.
$Sc^-$ is a measure of the non-reciprocity and can have a positive or negative sign.

In the following we will derive the Boltzmann equation with exact collision operator.
The Fokker-Planck equation for the $N$-particle probability distribution\\ $P_N(\vec{r}_1, ...,\vec{r}_N,\theta_1, ..., \theta_N,t)$ reads (with omitted arguments to improve readability)
\begin{eqnarray}
    \partial_t P_N = - \sum_i \left[\hat{n_i}\frac{\partial P_N}{\partial \vec{r_i}} - \frac{\partial}{\partial \theta_i}\sum_j a_{ij}Sc(\sigma_i, \sigma_j)\sin(\theta_j-\theta_i)P_N -\frac{\partial^2}{\partial\theta_i^2}\nu^2P_N\right].
    \label{eq:nonrec:liouville}
\end{eqnarray}
We obtain the first equation of the BBGKY hierarchy by integrating over the phases of all but the first particle, which we assume has coordinates $\vec{r},\theta$ and belongs to species $\sigma$.
We define the normalized probability densities $p_k={P_k}{L^{2k}}$. Then, in the thermodynamic limit where $N\gg 1$ and $L\gg 1$ at constant density
\begin{equation}
\begin{multlined}
        \partial_tp_1(\vec{r}, \theta| \sigma) = -\hat{n}(\theta)\cdot\frac{\partial p_1}{\partial \vec{r}} +\frac{\partial^2}{\partial\theta^2}\nu^2p_1\\* 
        - \sum_{\tilde{\sigma}}\rho(\tilde{\sigma})\frac{\partial}{\partial \theta}\int \mathrm{d}\beta \int \mathrm{d}\vec{z}\, a(\vec{r}-\vec{z})Sc(\sigma,\tilde{\sigma})\sin(\beta-\theta)p_2(\vec{r}, \vec{z}, \theta, \beta| \sigma, \tilde{\sigma})
        \label{eq:nonrec:hi1}
    \end{multlined}
    \end{equation}
We used the fact that $P_N$ is invariant under the permutation of the coordinates of particles from the same species. The notation $(...|\sigma)$ is used to keep track of the species of the particles involved.
The second equation of the hierarchy is obtained by integrating over the phases of all but two particles
\begin{equation}
\begin{multlined}
    \partial_tp_2(\vec{r}, \vec{z}, \theta, \beta| \sigma, \tilde{\sigma}) = -\hat{n}(\theta)\cdot\frac{\partial p_2}{\partial \vec{r}}-\hat{n}(\beta)\cdot\frac{\partial p_2}{\partial \vec{z}} +\frac{\partial^2}{\partial\theta^2}\nu^2p_2+\frac{\partial^2}{\partial\beta^2}\nu^2p_2 \\*
    + \rho(\tilde{\sigma})\sum_{\sigma_3}\rho(\sigma_3)\frac{\partial}{\partial \theta}\int \mathrm{d}\alpha \int \mathrm{d}\vec{y}\, a(\vec{r}-\vec{r_3})Sc(\sigma, \sigma_3)\sin(\theta_3-\theta)p_3(\vec{r}, \vec{z}, \vec{y}, \theta, \beta, \alpha| \sigma, \tilde{\sigma}, \sigma_3)\\* 
    + \rho(\tilde{\sigma})\sum_{\sigma_3}\rho(\sigma_3)\frac{\partial}{\partial \beta}\int \mathrm{d}\alpha \int \mathrm{d}\vec{y}\, a(\vec{z}-\vec{r_3})Sc(\tilde{\sigma}, \sigma_3)\sin(\theta_3-\beta)p_3(\vec{r}, \vec{z}, \vec{y}, \theta, \beta, \alpha| \sigma, \tilde{\sigma}, \sigma_3)\\*
    + \frac{\partial}{\partial \theta} a(\vec{r}-\vec{z})Sc(\sigma, \tilde{\sigma})\sin(\beta-\theta)p_2 + \frac{\partial}{\partial \beta} a(\vec{r}-\vec{z})Sc(\tilde{\sigma}, \sigma)\sin(\theta-\beta)p_2
    \label{eq:nonrec:hi2}
\end{multlined}
\end{equation}

The Boltzmann equation is derived by properly inserting \eqref{eq:nonrec:hi2} into the collision integral of \eqref{eq:nonrec:hi1} (see Refs. \cite{ihle2023asymptotically,ihle2023scattering} for details). 
Analogous to the derivation of the Boltzmann equation for regular, passive gases, see Ref. \cite{kreuzer1981nonequilibrium}, the terms containing $p_3$ can be neglected for low densities, where the probability of three particles being close together is small. 
This effectively closes the infinite BBGKY-hierarchy at the second level. An alternative, easier closure of the hierarchy at this level  
has been utilized recently in Ref. \cite{boltz2024kinetic}.
The term $\partial_t p_2=0$ can also be neglected on the short timescale of a collision, as detailed in Ref. \cite{ihle2023asymptotically}.
Furthermore, we assume that the noise is so small, i.e. $\nu^2\ll\min(1, Sc)$, that it does not significantly contribute on the short timescale of a collision.
The resulting equation therefore reads
\begin{equation}
    \begin{aligned}
        &\partial_tp_1(\vec{r}, \theta| \sigma) = -\hat{n}(\theta)\cdot\frac{\partial p_1}{\partial \vec{r}} +\frac{\partial^2}{\partial\theta^2}\nu^2p_1  +\sum_{\tilde{\sigma}}\rho(\tilde{\sigma})J_\mathrm{coll}^{\sigma\tilde{\sigma}},
        \label{eq:nonrecmain:closed}
    \end{aligned}
    \end{equation}
where $p_1(\vec{r}, \theta| \sigma)$ is the reduced density function for the species $\sigma$.
There is a collision integral for each different combination of species.
\begin{equation}
    \begin{aligned}
    J_\mathrm{coll}^{\sigma\tilde{\sigma}} &= \int_{-\pi}^{\pi}\mathrm{d}\Delta\ \int_{-\pi/2}^{\pi/2}\mathrm{d}\varphi\ \cos\varphi\ v_\mathrm{rel}\ p_2(\vec{r}, \vec{z}, \theta,\beta, t|\sigma,\tilde{\sigma}) \\*
    &+\int_{-\pi}^{\pi}\mathrm{d}\Delta\ \int_{\pi/2}^{3\pi/2}\mathrm{d}\varphi\ \cos\varphi\ v_\mathrm{rel}\ p_2(\vec{r}, \vec{z}, \theta,\beta, t|\sigma,\tilde{\sigma})\\*
    &=J_\mathrm{rec}^{\sigma\tilde{\sigma}}+J_\mathrm{app}^{\sigma\tilde{\sigma}},
    \end{aligned}
    \label{eq:nonrecmain:result1}
\end{equation}
where the second integration is along the contour of the collision
circle of the focal particle.
The variables of integration are chosen so that $\varphi$ is the polar angle between $\vec{v}_\mathrm{rel}$ and $\Delta\vec{r}=\vec{z}-\vec{r}$, and $\Delta=\beta-\theta$.
The first term in \eqref{eq:nonrecmain:result1} describes configurations at the end of a collision and is labelled the receding part.
Conversely the second term includes configurations at the start of a collision and is labelled the approaching part.
A closed equation for $p_1$ is obtained by employing the one-sided molecular chaos approximation, meaning that particles are assumed to be uncorrelated before a collision but not afterwards.
This can be done right away in the approaching part of the collision integral, but in the receding integral it requires expressing $p_2$ in terms of its value at the start of the collision, where a factorization in terms of $p_1$ is sensible.
This backtracking to the beginning of the collision is achieved by solving the equations of motion for a collision of two particles, neglecting noise as well as encounters of three or more particles.
While up to this point the results could be straightforwardly transferred from the reciprocal system, here the differences will become apparent as solving the equations of motion is more complicated for non-reciprocal interactions.
Due to the non-Hamiltonian equations of motion, we have
\begin{equation}
    \frac{dp_2}{dt}=\Lambda\,p_2
\end{equation}
where $\Lambda$ is called the phase-space compression factor \cite{j2007statistical}.
In this specific case $\Lambda=-Sc^+(\sigma,\tilde{\sigma}) \cos(\beta-\theta)$ and therefore 
\begin{equation}
p_2(\vec{r}(t), \vec{z}(t), \theta(t), \beta(t),t)=p_2(\vec{r}(t_0), \vec{z}(t_0), \theta(t_0), \beta(t_0),t_0)\exp\left(-Sc^+\int_{t_0}^t \cos\Delta(t')dt'\right),
\end{equation}
where $t_0$ is the time the collision started.
This formula is the same as in the reciprocal case, except for the replacement $2Sc \rightarrow Sc^+$.
The same is true for the time evolution of $\Delta$
\begin{equation}
    \tan(\Delta(\bar{t})/2)=\mu\exp(-Sc^+(t-\bar{t})).
    \label{eq:nonrec:angle}
\end{equation}
where $\mu=\tan(\Delta(t)/2)$.
The directional angles before the collision are obtained using the formulas for their linear combinations $\Delta$ and $s=\theta+\beta$.
While $s$ conveniently stays constant during a reciprocal interaction, in the general case it is given by
\begin{equation}
    s(\bar{t})=s(t)-Sc^-\int_t^{\bar{t}}\sin\Delta(t')\ dt'= s(t)-\frac{2Sc^-}{Sc^+}\left[\arctan(\mu\exp(-Sc^+(t-\bar{t})))-\arctan(\mu)\right].
    \label{eq:nonrec:sum}
\end{equation}
To evaluate equations \eqref{eq:nonrec:angle} to \eqref{eq:nonrec:sum} at the start of a collision $t_0$, we need to know the duration of a collision $t_{dur}=t-t_0$.
It is obtained by solving the equations of motion for the positions of two particles.
This cannot be done exactly for nonreciprocal interactions, but it is possible perturbatively for small $Sc$.
We therefore try the ansatz that $t_{dur}$ is of similar form as for reciprocal interactions, with a small correction $\delta$
\begin{equation}
    t_{dur}= -\frac{1}{Sc^+}\ln\left\{\frac{1}{\mu}\sinh\left[-Sc^+\cos(\varphi)\text{sgn}(\sin(\Delta/2))+\arcsinh\mu+\delta\right]\right\},
    \label{eq:nonrec:tdur}
\end{equation}
which leads to the following expression for $\delta$
\begin{equation}
    \delta=-\frac{1}{2}Sc^-Sc^+\cos\frac{\Delta}{2}\sin\varphi\cos\varphi+O(Sc^3).
\end{equation}
One can show that up to second order in $Sc$ this correction does not contribute, since it is uneven in $\varphi$ and therefore vanishes in the collision integral.
The expression \eqref{eq:nonrec:tdur} diverges if the argument of the $\ln$-function is negative.
This necessitates the definition of a \textit{forbidden zone} of post-collisional configurations that cannot be reached through a natural evolution of the system.
This zone must be excluded in the evaluation of the receding collision integral.
The mathematical treatment of this zone can be found in the appendix, here we only note that the higher order correction term $\delta$ also does not contribute to the forbidden zone up to the required order of accuracy.
If we neglect this correction to $t_{dur}$, it is again the same as in the reciprocal case with the replacement $2Sc \rightarrow Sc^+$.
The only additional terms in the nonreciprocal collision integral therefore come from Eq. \eqref{eq:nonrec:sum}.
The collision integral is transformed to Fourier space
\begin{subequations}
    \begin{eqnarray}
        \hat{p}_n(\vec{r}, t) = \int_0^{2\pi}\mathrm{d}\theta\ p_1(\vec{r}, \theta, t)\exp(-\mathrm{i} n \theta)\\*
        p_1(\vec{r}, \theta, t) = \frac{1}{2\pi}\sum_{n=-\infty}^{\infty}\hat{p}_n(\vec{r}, t) \exp(\mathrm{i} n \theta).
    \end{eqnarray}
    \label{eq:kinetic:Fourier}
\end{subequations}
The mathematical details of solving the two particle dynamics and the collision integral can be found in the appendix. 
Here we only state the result
\begin{equation}
    \begin{aligned}
        \hat{J}_{\mathrm{coll}, m}^{\sigma,\tilde{\sigma}} &= -Sc(\sigma,\tilde{\sigma}) \frac{m\pi}{2}\left(\hat{p}_{m-1}(\sigma)\hat{p}_{1}(\tilde{\sigma})-\hat{p}_{m+1}(\sigma)\hat{p}_{-1}(\tilde{\sigma})\right)\\*
        & + \left(\frac{Sc^+(\sigma,\tilde{\sigma})}{2}\right)^2\sum_n \hat{p}_{m-n}(\sigma)\hat{p}_{n}(\tilde{\sigma})g_{mn}\\*
        & + \left(\frac{Sc^+(\sigma,\tilde{\sigma})}{2}\right)Sc^-(\sigma,\tilde{\sigma})\sum_n \hat{p}_{m-n}(\sigma)\hat{p}_{n}(\tilde{\sigma})h_{mn}\\*
        & + Sc^-(\sigma,\tilde{\sigma})^2\sum_n \hat{p}_{m-n}(\sigma)\hat{p}_{n}(\tilde{\sigma})f_{mn}
        \label{eq:nonrecmain:final}
    \end{aligned}
    \end{equation}
with the coupling matrices
\begin{equation}
    \begin{aligned}
        g_{mn} = \frac{1}{2\pi}\frac{8}{3}m\left(\frac{3/2m-n}{n^2-9/4}+\frac{1/2m+n}{n^2-1/4}\right)
    \end{aligned}
    \end{equation}
\begin{equation}
    \begin{aligned}
        h_{mn} = \frac{1}{2\pi}\frac{64}{3}m\left(\frac{3m+3n-4mn^2}{9-40n^2+16n^4}\right)
    \end{aligned}
    \end{equation}
    \begin{equation}
        \begin{aligned}
            f_{mn} = \frac{1}{2\pi}\frac{16}{3}m^2\left(\frac{4n^2-3}{9-40n^2+16n^4}\right).
        \end{aligned}
        \end{equation}
The first order term in Eq. \eqref{eq:nonrecmain:final} is again equivalent to the mean field equation.
The second order term contains one contribution that is equivalent to the reciprocal case with averaged coupling strengths (the term containing $g_{mn}$ \cite{ihle2023asymptotically,ihle2023scattering,kursten2023flocking}).
The remaining two terms are only present for non-reciprocal interactions, and are one of the main results of this paper.

\section{Comparison with agent-based simulations}
\begin{figure}
    \includegraphics[width=.45\textwidth]{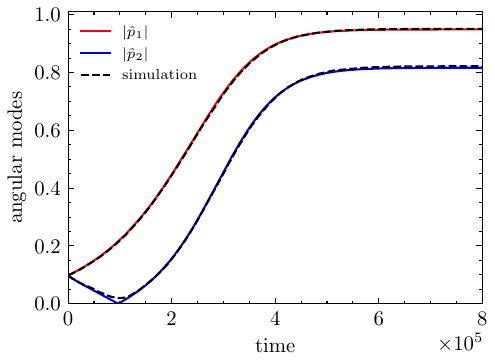}
    \includegraphics[width=.45\textwidth]{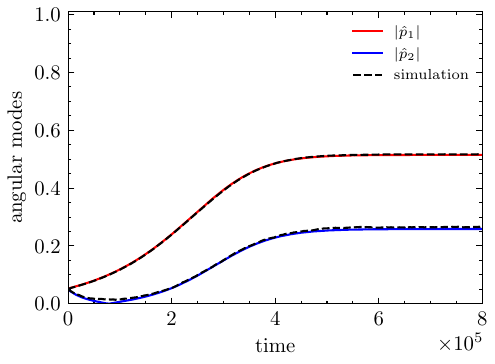}
\caption{Time evolution of the first two Fourier modes of the two species obtained from the theory (red and blue solid lines) in comparison with agent-based simulations (black dashed lines). Species 1 is shown on the left, species 2 on the right. Parameters are $M_1=0.03$, $M_2=0.06$ and $Sc(1,1)=0.01$, $Sc(2,2)=0.018$, $Sc(1,2)=0.01$, $Sc(2,1)=0.02$. Here, and in the following figures, no external noise was included, $\nu=0$.
The difference between simulation and theory when the modes are very close to zero is an artifact of the numerical evaluation of the simulation, where it was more efficient to only measure the absolute value of the modes.}
\label{fig:nonrec:modes}
\end{figure}

We compare the results to agent-based simulations in a binary system containing species 1 and 2.
All simulations are performed using the AAPPP Python package \cite{kursten2023aligning}, with a timestep $\Delta t=0.02$ and in a quadratic domain with $L=560.5$.
We first measure the time evolution of the angular distributions $p_n$. 
The system size is small enough that we can assume spatial homogeneity, preventing the formation of long-wavelength instabilities of ordered states \cite{kursten2023flocking,bertin2006boltzmann,bertin2009hydrodynamic,chate2008collective,romanczuk2012active,ihle2013invasion}.
We therefore evaluate the analytical results by numerically integrating the following mode hierarchy 
\begin{equation}
    \partial_t\hat{p}_m(\vec{r}, t|\sigma) = \hat{J}_{\mathrm{coll}, m}^{\sigma\sigma} + \hat{J}_{\mathrm{coll}, m}^{\sigma\bar{\sigma}}
\label{eq:nonrecmain:hi}
\end{equation}
assuming modes $|m|>50$ are zero. We take an ensemble average over 10 realizations of the system.
As initial conditions, the positions of the particles are distributed uniformly over the whole computational domain.
The directions of species 1 are uniformly distributed on the interval $\left[-a,a\right]$, that of species 2 on the interval $\left[-b+\pi,b+\pi\right]$.
This means the flying directions of species 1 are in a sector of a circle pointing to the left, those of species 2 in a sector pointing to the right.
The corresponding Fourier modes are $\hat{p}_n(1)=\sinc(na)$ and $\hat{p}_n(2)=-\sinc(nb)$.
We choose $a=2.85234$ and $b=2.99146$ because that amounts to $\hat{p}_1(1)=0.1$ and $\hat{p}_1(2)=-0.05$.
Figure \ref{fig:nonrec:modes} shows that there is excellent quantitative agreement, both in the time evolution and for the stationary values of the
Fourier modes.

\begin{figure}
        \centering
    \includegraphics[width=.245\textwidth]{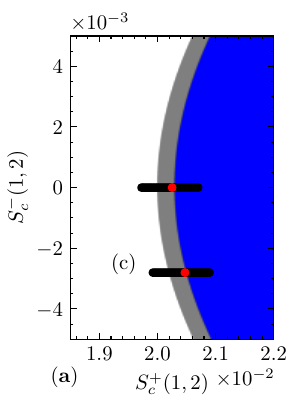}
    \hspace*{-15pt}
    \includegraphics[width=.245\textwidth]{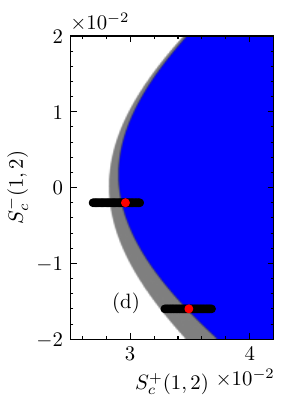}
    \hspace*{+10pt}
    \includegraphics[width=.245\textwidth]{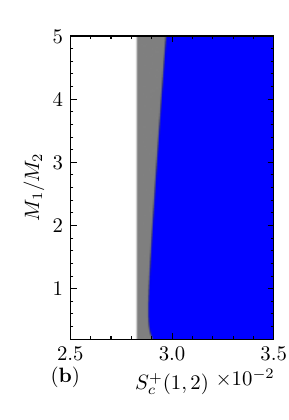}
    \hspace*{-15pt}
    \includegraphics[width=.245\textwidth]{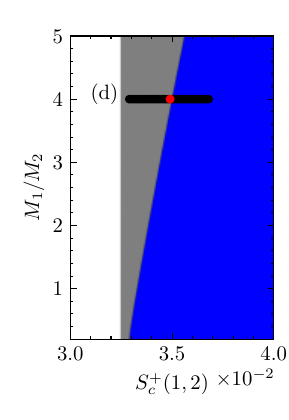}

    \includegraphics[width=.49\textwidth]{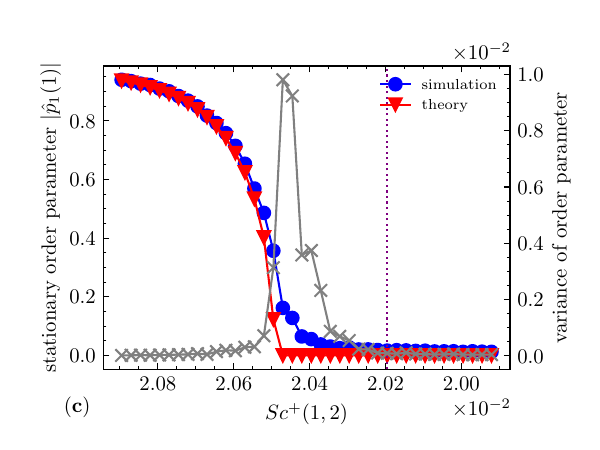}
    \includegraphics[width=.49\textwidth]{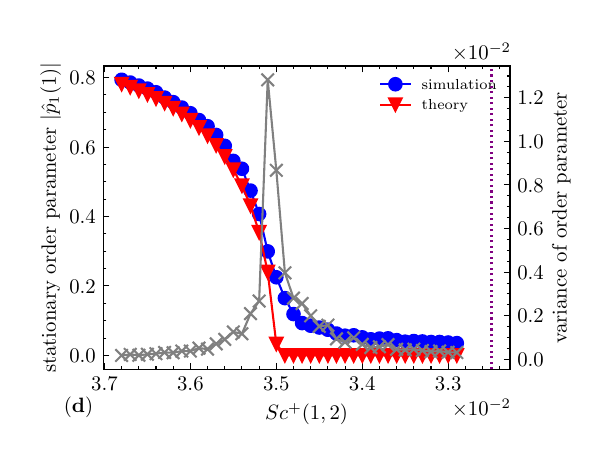}
\caption{\linespread{1.26}\selectfont{} (a) and (b) show where the disordered solution of the Boltzmann equation (blue) and the mean field equation (gray+blue) becomes linearly unstable.
Figure (a) illustrates the influence of non-reciprocity on the location of the transition. All other parameters are symmetric between the two species in the left frame ($Sc(1,1)=Sc(2,2)=0.01$, $M_1=M_2=0.03$). In the right frame the species also differ in intra-species-coupling strength and density, ($2Sc(1,1)=Sc(2,2)=0.01$, $4M_1=M_2=0.03$).
Figure (b) illustrates the influence of the density ratio $M_1/M_2$ on the location of the transition. In the left frame of (b), the interactions are reciprocal ($2Sc(1,1)=Sc(2,2)=0.01$, $Sc^-(1,2)=0$), in the right frame they are non-reciprocal ($2Sc(1,1)=Sc(2,2)=0.01$, $Sc^-(1,2)=-0.016$).
Along the short black lines, agent-based simulations were performed for comparison. The red dot shows where the transition to an ordered stationary state takes place in the agent-based simulation.
(c) and (d) show the group of simulations marked in (a) and (b). Blue dots are the stationary order parameters $|\hat{p}_1(1)|$ measured in the simulation, whereas red triangles represent the stationary values predicted by the theory. Grey crosses show the variance of the order parameter measured in simulation. The maximum of this variance was used to determine the exact location of the transition.
The purple dotted line is the mean field stability line.
The simulations were performed along lines with constant $Sc^-$ ($Sc^-=-0.00280435$ in (c) and $Sc^-=-0.016$ in (d)).
}
\label{fig:nonrec:overview}
\end{figure}

The location of the transition to a flocking state in a small spatially homogeneous system can be found by linear stability analysis of the disordered state.
Eq.~\eqref{eq:nonrecmain:hi} reads in matrix form
\begin{equation}
    \partial_t \begin{pmatrix}\hat{p}_1(1)\\* \hat{p}_1(2)\end{pmatrix}=\mathcal{M} \begin{pmatrix}\hat{p}_1(1)\\* \hat{p}_1(2)\end{pmatrix}
\end{equation}
with the diagonal elements
\begin{equation}
\begin{multlined}
    \mathcal{M}_{\sigma,\sigma} = \rho(\sigma)\left(-Sc(\sigma,\sigma)\frac{m\pi}{2} +Sc^2(\sigma,\sigma)(g_{1,0}+g_{1,1})\right)\\* + \rho(\bar{\sigma})\left(\left(\frac{Sc^+(\sigma,\bar{\sigma})}{2}\right)^2g_{1,0}+\frac{Sc^+(\sigma,\bar{\sigma})}{2}Sc^-(\sigma,\bar{\sigma})h_{1,0}+Sc^-(\sigma,\bar{\sigma})^2f_{1,0}\right)
\end{multlined}
\end{equation}
and the off diagonal elements
\begin{equation}
\begin{multlined}
    \mathcal{M}_{\sigma,\bar{\sigma}} = \rho(\bar{\sigma})\left(-Sc(\sigma,\bar{\sigma})\frac{m\pi}{2} +\left(\frac{Sc^+(\sigma,\bar{\sigma})}{2}\right)^2g_{1,1}\right.\\* \left.+\frac{Sc^+(\sigma,\bar{\sigma})}{2}Sc^-(\sigma,\bar{\sigma})h_{1,1}+Sc^-(\sigma,\bar{\sigma})^2f_{1,1}\right).
\end{multlined}
\end{equation}
As the trace of $\mathcal{M}$ is negative, it can only have a positive eigenvalue if $\det(\mathcal{M})<0$.
Therefore the condition to observe a flocking state is 
\begin{equation}
    \mathcal{M}_{1,1}\mathcal{M}_{2,2}<\mathcal{M}_{1,2}\mathcal{M}_{2,1}.
    \label{eq:nonrecmain:flocking}
\end{equation}
When keeping only leading order terms (equivalent to mean field theory) this condition becomes
\begin{equation}
    Sc(1,1)Sc(2,2)<Sc(1,2)Sc(2,1).
    \label{eq:nonrecmain:flockingmf}
\end{equation}
Figure \ref{fig:nonrec:overview} shows the domain of the parameter space where flocking is predicted.
This prediction is validated by integrating the mode equations Eq.~\eqref{eq:nonrecmain:hi} until a stationary state is reached (this was determined visually) and then averaging over the following $1.5\cdot 10^{8}$ iterations.
The first Fourier mode of the first species $|\hat{p}_1(1)|$ was used as the polar order parameter to quantify possible
order-disorder transitions.
The general trend, shown in figure \ref{fig:nonrec:overview}(a), is that increasing the non-reciprocity $|Sc^-|$ moves the transition to flocking to higher values of inter-species anti-alignment $Sc^+$.
An important difference to the mean-field theory is that the transition is now dependent on the ratio of the particle numbers $M_1/M_2$.
In figure \ref{fig:nonrec:overview}(b) one can see that the influence of this ratio is strengthened by the non-reciprocity.

\begin{figure}
    \centering
    \includegraphics[width=.5\textwidth]{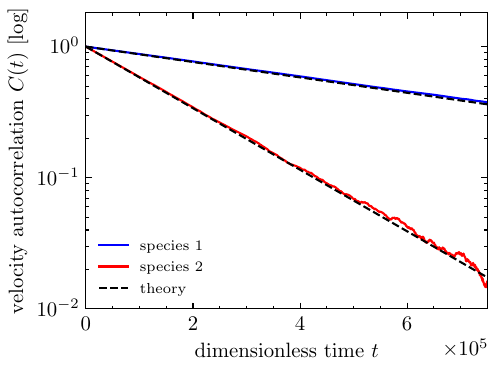}
    \caption{Velocity autocorrelation of species 1 (blue) and species 2 (red). Solid colored lines refer to agent-based simulations, while the dashed black lines show the predictions from Eq.~\eqref{eq:nonrec:tau}. Parameters are $M_1=M_2=0.03$ and  $Sc(1,1)=Sc(2,2)=0.01$, $Sc(1,2)=0.005$, $Sc(2,1)=0.002$.}
    \label{fig:autocorrelation}
\end{figure}

In the disordered phase, the theory can also predict the self-diffusion coefficient. We proceed analogous to Ref. \cite{ihle2023asymptotically}, by formally introducing a third species that consists only of one tagged particle of species $\sigma$.
In the disordered state the density function $h(\vec{r},\theta)$ of this third species therefore follows as
\begin{equation}
\begin{multlined}
    \partial_t \hat{h}_1 = \hat{h}_1\rho(\sigma) Sc(\sigma,\sigma)^2g_{1,0} 
    +\hat{h}_1\rho(2)\left[\left(\frac{Sc^+(\sigma,\bar{\sigma})}{2}\right)^2 g_{1,0}\right. \\*\left. +\left(\frac{Sc^+(\sigma,\bar{\sigma})}{2}\right)\left(Sc^-(\sigma,\bar{\sigma})\right) h_{1,0} +\left(Sc^-(\sigma,\bar{\sigma})\right)^2 f_{1,0}\right].
\end{multlined}
\end{equation}
The function $\hat{h}_1$ can be shown to be proportional to the velocity autocorrelation, and therefore we obtain for the correlation time $\tau$
\begin{equation}
\begin{multlined}
    \tau = - \left[\rho(\sigma) Sc(\sigma,\sigma)^2g_{1,0}
    +\rho(2)\left(\left(\frac{Sc^+(\sigma,\bar{\sigma})}{2}\right)^2 g_{1,0}\right.\right.\\*\left.\left. +\left(\frac{Sc^+(\sigma,\bar{\sigma})}{2}\right)\left(Sc^-(\sigma,\bar{\sigma})\right) h_{1,0} +\left(Sc^-(\sigma,\bar{\sigma})\right)^2 f_{1,0}\right)\right]^{-1}.
\end{multlined}
\label{eq:nonrec:tau}
\end{equation}
By means of the usual Green-Kubo relation, the self-diffusion 
coefficient $D$ follows from the relaxation time $\tau$ as
\begin{equation}
    D=\frac{\tau}{2},
    \label{eq:kinetic:diffcoeff}
\end{equation}
see Ref. \cite{ihle2023asymptotically}.
The autocorrelation was measured in agent-based simulations as an average over all particles and 10 independent realizations.
The results are shown in figure \ref{fig:autocorrelation} and agree very well with the theory.

\begin{figure}
    \includegraphics[width=.49\textwidth]{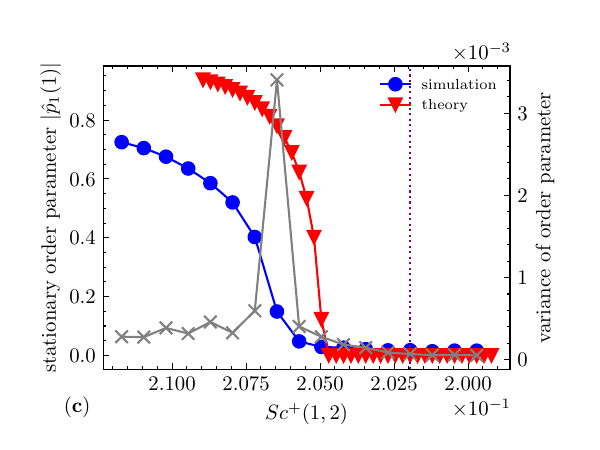}
    \caption{ A version of Fig. \ref{fig:nonrec:overview}(c) with the same relative values of coupling and density, but higher absolute values: $Sc(1,1)=Sc(2,2)=0.1$, $Sc^-=-0.0280435$, and $M_1=M_2=15.62$.
Blue dots are the stationary order parameters $|\hat{p}_1(1)|$ measured in the simulation, whereas red triangles represent the stationary values predicted by the theory. Grey crosses show the variance of the order parameter measured in simulation. The maximum of this variance was used to determine the exact location of the transition.
The purple dotted line is the mean field stability line.}
    \label{fig:dishighdens}
\end{figure}

Generally, we expect our active Boltzmann theory to give accurate quantitative predictions for small densities $M\ll 1$ and small coupling $Sc\ll1$.
The requirement of small densities can be relaxed if the coupling is sufficiently small, $MSc\ll1$.
This is because at small couplings the Boltzmann equation
usually agrees with the Landau equation, and the Landau approach can be shown diagrammatically to be valid for $MSc\ll 1$,
see Ref.~\cite{boltz2024kinetic}.
We give an example of the deviation of the analytic predictions from the simulations at higher densities in Fig. \ref{fig:dishighdens}.
In this figure the parameters have the same values relative to each other as in Fig. \ref{fig:nonrec:overview}(c), but higher absolute values.
With $M_1=M_2=15.6$, the density is much larger than one, and $MSc\approx 1$.
The location of the transition to polar order is still predicted very well, with a difference of only 1\%, but the stationary value of the Fourier modes is significantly overestimated by the theory.

The sum of all flying angles $S$ is complicated to derive directly from the kinetic equation, as it is not a $2\pi$-periodic quantity.
We will instead use a simple argument which includes the assumption of binary collisions, which is also central to the Boltzmann equation.
For this argument we assume for simplicity that the system is in a non-flocking, unordered state.
We can again approximate the dynamics as a sequence of independent two-particle collisions.
For one such collision, we can solve the equations of motion perturbatively for small coupling and calculate the squared change of the sum of flying angles $\delta s^2(\Delta,\varphi)=4\cos^2(\Delta/2)\cos^2(\varphi) (Sc^-)^2$ (see appendix \ref{app:add} for details).
During a time interval of length $t$, each of the $N$ particles sweeps an area of $tv_{rel}(\Delta)\sigma(\varphi)\mathrm{d}\varphi$ relative to a second particle characterized by $\Delta$ and $\varphi$.
Here $v_{rel}(\Delta)=2v_0|\sin(\Delta/2)|$ is the relative velocity of the two particles, and $\sigma(\varphi)=R\cos(\varphi)$ is the differential cross-section.
Per unit area there are a number of $N/L^2\cdot\mathrm{d}\Delta/(2\pi)$ particles with relative flying direction $\Delta$.
Thus the frequency of mixed-species collisions is 
\begin{equation}
f_\mathrm{coll}(\Delta,\varphi)\mathrm{d}\Delta\mathrm{d}\varphi=\frac{N_1N_2}{N^2}\frac{tv_{rel}(\Delta)\sigma(\varphi)N\frac{N}{L^2}\frac{\mathrm{d}\Delta}{2\pi}\mathrm{d}\varphi}{t} = \frac{Rv_0}{\pi}|\sin(\Delta/2)|\cos(\varphi)\frac{N_1N_2}{L^2},
\end{equation}
where the combinatorical factor $\frac{N_1N_2}{N^2}$ accounts for the fact that only a fraction of collisions involves two different species.
Between collisions, $S$ only changes due to the stochastic noise term, therefore $\dot{\langle S^2\rangle}=\sum_j\dot{\langle \theta_j^2\rangle}=2N\nu^2$.
We thus arrive at the expression
\begin{equation}
    \frac{\mathrm{d}}{\mathrm{d}t}\langle S^2\rangle=\int_{-\pi/2}^{\pi/2}\mathrm{d}\varphi\int_{-\pi}^{\pi}\mathrm{d}\Delta\ f_\mathrm{coll}(\Delta,\varphi)\delta s^2(\Delta,\varphi)+2N\nu^2=\frac{64}{9\pi}\frac{N_1N_2}{L^2}(S_c^-)^2+2N\nu^2,
    \label{eq:deltaSsqu}
\end{equation}
which agrees well with simulations, as shown in figure \ref{fig:sum}. This last equation is again in units where $R=v_0=1$.

\section{Conclusions}
We investigated binary mixtures of self-propelled particles with non-reciprocal
anti-aligning interactions.
The dynamics is governed by Kuramoto-style Langevin-equations where
neighboring particles exert torques on each other.
There is a hidden variable in these systems, the sum of the flying angles of all
particles, whose conservation is broken if the torques are non-reciprocal.
The most common kinetic theories for systems with alignment interactions are of the mean-field type,
see e.g. Fruchart et al. \cite{fruchart2021non}, Menzel \cite{menzel2012collective}, Peruani et al. \cite{peruani2008mean}, Lee \cite{lee2010fluctuation}
However, mean-field theory leads to quantitatively and sometimes even qualitativly incorrect predictions
such as an infinite diffusion
coefficient \cite{ihle2023scattering,ihle2023asymptotically} or the absence of a flocking transition in particular models of polar active
particles \cite{das2024flocking,ketzetzi2024self}.
Because of this and since the implications of the above mentioned symmetry breaking are not clear,
we go beyond mean-field and use an active Boltzmann approach with exact collision operator.
This approach is based on a non-local closure of the BBGKY hierarchy in the limit of small densities.
For this it is necessary to solve the equations of motion analytically for a collision of two particles, something that is more complicated for the non-reciprocal interactions considered here compared to the reciprocal interactions investigated with this approach previously.
Because of that we performed our calculations perturbatively for small coupling.
We derive equations for the time evolution of the one-particle distribution, as well as an expression for the coefficient of self-diffusion.
We found excellent quantitative agreement
between agent-based simulations and
the predictions of the active Boltzmann equation not only for steady but also for
transient states in parameter ranges where the product of density M and coupling strength Sc is small.
While the condition for polar order depends only on the coupling constants in mean field theory Eq.~\eqref{eq:nonrecmain:flockingmf}, the higher order terms obtained from the Boltzmann theory Eq.~\eqref{eq:nonrecmain:flocking} also introduce a dependence on the density and the velocity of the particles.
Qualitatively, this is all in accordance with the results from Ref. \cite{kursten2023flocking} for the reciprocal case.

In the reciprocal system, it was also found by means of agent-based simulations and mean-field theory that the spatially homogeneous solution in the flocking state is linearly unstable for large wavelength perturbations \cite{kursten2023flocking}.
As a consequence, microphase separation is observed.
Two different patterns form.
The first are stripes with high polar order alternating with stripes of no polar order, where the density of one species is larger in the ordered stripes.
The second pattern are patches with different polarization direction.
A recent study investigating reciprocal interactions at higher densities also found patterns of polarized stripes \cite{lardet2025flocking}.
We expect these effects in the non-reciprocal case as well.
The data we obtained from agent-based simulations is for small systems, where the assumption of spatial homogeneity is justified. In larger systems, the picture will look different.
The section above should therefore not be understood as a characterization of the phase diagram of the binary non-reciprocal system.
This is left for future research.
Here, we have shown that the Boltzmann equation can also quantitatively describe non-reciprocal interactions and that accurate analytical predictions for transport coefficients
can be made.
As the Boltzmann equation contains many more terms than the mean field equation, solving for the stationary solution in a binary system is not possible analytically.
Investigating spatially inhomogeneous solutions in a binary system in the framework of the Boltzmann equation is therefore best done by solving the full equation numerically, similar to Refs. \cite{ihle2013invasion,thuroff2014numerical}.
To describe qualitatively new effects that are caused specifically by the non-reciprocal interactions, it would be interesting to extend the Boltzmann approach to strong polar aligning interactions in the future, something that is beyond the reach of
alternative weak coupling approaches such as Ref. \cite{boltz2024kinetic}.
Our model is based on point-like particles. A possible extension is to include hard-core repulsion between particles.
As long as the range of this repulsion $r_\mathrm{rep}$ is small, $r_\mathrm{rep}\ll R$, we expect our results to still hold.
The reason is that the anti-alignment interactions will lead to the particles turning away from each other before they collide.
Those collisions between the hard cores of non-point-like particles will predominantly occur in the flocking state of two species with opposite average flying directions.
At higher densities, the excluded volume effect might therefore prevent the emergence of an ordered state.
At this point, there could also be a MIPS-transition under the influence of anti-alignment.
This interplay could be an interesting subject of future research.
The recent interest in  active matter with non-reciprocal aligning interactions comes from the fact that time dependent stationary solutions can be observed in the flocking state in certain parameter ranges \cite{fruchart2021non,kreienkamp2022clustering}.
This requires however that the two species have opposite goals (e.g. $Sc(1,2)\approx -Sc(2,1)$), leading to a sort of chasing motion.
In the system investigated in this study however, where all interactions are anti-aligning, we require $Sc^+=Sc(1,2)+Sc(2,1)\gg 0$ in order to observe a flocking state.
As those conditions cannot be fulfilled simultaneously, we do not observe these time-dependent states here.
To summarize, in this paper we have demonstrated that the active Boltzmann approach, originally introduced in \cite{ihle2023scattering,ihle2023asymptotically},
can be generalized to more complicated orientational interactions as in the simple one-component Vicsek-model, still leading
to analytical expressions and predictions, beyond the mean-field results of Fruchart et al. \cite{fruchart2021non}. We hope that in the future this theory will be useful to quantitatively
describe phenomena in
other interesting active matter systems that are experimentally relevant such as the
``flocking by turning away'' effect in Ref. \cite{das2024flocking} or self-reconfiguring active matter, Ref. \cite{ketzetzi2024self}.

\section*{Acknowledgement}
We thank H-H. Boltz for valuable discussions. We are thankful to the computer center of the 
University of Greifswald for computational time used in parts of this work.

\bibliography{cite}

\appendix

\section{Solving the collision integral for small reciprocal coupling}
\label{app:smallsc}
Here we follow Ihle et al. \cite{ihle2023asymptotically} and evaluate the collision integral of the Boltzmann equation in an expansion for small coupling $Sc(i,j)=Sc(j,i)=:Sc$.
This section is basically a recap of Ref. \cite{ihle2023asymptotically,ihle2023scattering} for the sake of selfcontainedness. However we change some definitions and the way the forbidden zone is treated mathematically.
This makes it much easier to adapt the calculations later on to the non-reciprocal case.
\subsection{Two particle dynamics}
To solve the equations of motion for a collision of two particles, we first make some useful definitions.
We have already defined 
\begin{equation}
    \Delta = \beta-\theta
    \label{eq:derivation:delta}
\end{equation}
in the main text. 
Furthermore, we define the sum of the angles
\begin{equation}
    \bar{c}=\theta+\beta
    \label{eq:derivation:cbar}
\end{equation}
By calculating the derivative of $\bar{c}$ via the microscopic rules Eq.~\eqref{eq:kinetic:dimensionlessrules}, one can see that it is constant during a collision.
Another constant that comes up often is
\begin{equation}
    c = \bar{c}-2\varphi
    \label{eq:derivation:c}
\end{equation}
This is a constant \textit{by definition}, as $\varphi$ only refers to the angle at the end of the collision.
The following identities can be shown (see Appendix \ref{app:identnonrec})
\begin{equation}
    v_\mathrm{rel} = 2 \sin(\Delta/2)
    \label{eq:derivation:vrel}
\end{equation}
\begin{equation}
    \sin(c/2) = -\cos(\varphi)\text{sgn}(\sin(\Delta/2))
    \label{eq:derivation:sinhalfc}
\end{equation}
With these definitions and identities, we can now turn to solving the equations of motion. 
For the difference of the flying directions, one finds,
\begin{equation}
    \dot{\Delta} = \dot{\beta}-\dot{\theta} = -2Sc\sin(\theta-\beta)= 2Sc \sin(\Delta)
\end{equation}
Separation of the variables leads to 
\begin{equation}
    \tan(\Delta(t')/2)=\tan(\Delta(t)/2)\exp(-2Sc(t-t')).
    \label{eq:derivation:angle}
\end{equation}
In the following, we will sometimes use the abbreviation $\mu=\tan(\Delta(t)/2)$.
To calculate $t_0$, we also need the dynamics of the positions.
We get 
\begin{equation}
\begin{aligned}
    \frac{d\,\Delta\vec{r}}{dt} = \begin{pmatrix}
        \cos\beta-\cos\theta\\* \sin\beta-\sin\theta
    \end{pmatrix} = \begin{pmatrix}
        -2\sin(\bar{c}/2)\sin(\Delta/2)\\* 2\cos(\bar{c}/2)\sin(\Delta/2)
    \end{pmatrix}
\end{aligned}
\end{equation}
using trigonometric identities.
Integrating this equation leads to 
\begin{equation}
\begin{aligned}
    \Delta\vec{r}(t_0)-\vec{\Delta r}(t) = \begin{pmatrix}
        -2\sin(\bar{c}/2)\\* 2\cos(\bar{c}/2)
    \end{pmatrix}G
    \label{eq:derivation:position}
\end{aligned}
\end{equation}
where we defined
\begin{equation}
\begin{aligned}
    G=\int_t^{t_0}\sin(\Delta(t')/2)dt'=\frac{1}{-2Sc}\left(-\arcsinh(\mu\exp(-2Sc(t-t_0))+\arcsinh\mu)\right).
\label{eq:derivation:gint}
\end{aligned}
\end{equation}
Here we substituted Eq.~\eqref{eq:derivation:angle} to calculate the integral.
The next step is to bring $\vec{\Delta r}(t)$ to the other side of Eq.~\eqref{eq:derivation:position} and square the equation. We look at collisions where $\vec{\Delta r}(t)=\begin{pmatrix}
    \cos\varphi& \sin\varphi
\end{pmatrix}^T$ and where the collision began at $t_0$ with $\vec{\Delta r}(t_0)^2=1$. Therefore we end up with the simple expression
\begin{equation}
\begin{aligned}
    0=4G\left(G-\sin(c/2)\right)
\end{aligned}
\end{equation}
This has the trivial solution $t_{dur}=t-t_0=0$ and a non-trivial solution
\begin{equation}
    t_{dur}= \frac{1}{-2Sc}\ln\left\{\frac{1}{\mu}\sinh\left[2Sc\sin(c/2)+\arcsinh\mu\right]\right\}
    \label{eq:derivation:tdur}
\end{equation}

\subsection{Forbidden zone}
The duration $t_{dur}=t-t_0$ diverges (remembering identity Eq.~\eqref{eq:derivation:sinhalfc}) if
\begin{equation}
    -2Sc\cos\varphi+\arcsinh(\tan(|\Delta/2|))<0
    \label{eq:derivation:forbidden_condition}
\end{equation}
When this condition is met, the particles are still almost perfectly aligned after their collision ($\Delta$ is very small). This is something that should be prevented by the anti-aligning interactions of the particles.
We therefore conclude that these certain configurations do not occur at all in reality. This is called the "forbidden zone" where we must set $p_2=0$. We realize this by excluding those configurations from the integral. Therefore the integral of the receding part becomes
\begin{equation}
    \int_{-\pi/2}^{\pi/2}\mathrm{d}\varphi\ \left(\int_{-\pi}^{-\Delta_c}\mathrm{d}\Delta\ +\int_{\Delta_c}^{\pi}\mathrm{d}\Delta\ \right)
    \label{eq:derivation:forbidden}
\end{equation}
with 
\begin{equation}
    \Delta_c=2\arctan(\sinh(2Sc\cos\varphi)).
    \label{eq:derivation:Deltac}
\end{equation}
As $t_{dur}$ only occurs in exponential functions in the angular dynamics Eq.~\eqref{eq:derivation:angle} (and also in $J_s$, as we will see further below), the integral can still be defined in a sensible way inside the forbidden zone. Therefore, a formally equivalent formulation to Eq.~\eqref{eq:derivation:forbidden} is to integrate over the whole domain and subtract the integral over the part we want to cut out
\begin{equation}
    \int_{-\pi}^{\pi}\mathrm{d}\Delta\ \int_{-\pi/2}^{\pi/2}\mathrm{d}\varphi\ -\int_{-\Delta_c}^{\Delta_c}\mathrm{d}\Delta\ \int_{-\varphi_c}^{\varphi_c}\mathrm{d}\varphi
    \label{eq:derivation:forbidden_sub}
\end{equation}
So instead of Eq.~\eqref{eq:nonrecmain:result1}, we have
\begin{equation}
    J_\mathrm{coll} =J_\mathrm{rec}+J_\mathrm{app}+J_{for}
    \label{eq:derivation:result2}
\end{equation}
where
\begin{equation}
    J_{for}=-\int_{-\pi/2}^{\pi/2}\mathrm{d}\varphi\ \int_{-\Delta_c}^{\Delta_c}\mathrm{d}\Delta\ \cos\varphi\ v_\mathrm{rel}\ e^{-2 Sc J_s}\ p_1(\vec{r}(t_0), \theta(t_0), t_0)\ p_1(\vec{z}(t_0), \beta(t_0), t_0).
    \label{eq:derivation:jfor}
\end{equation}

\subsection{Approaching part}
What is left to do is evaluating the integrals in Fourier space according to Eq.~\eqref{eq:kinetic:Fourier}.
The Fourier-transformed approaching integral is
\begin{equation}
\begin{aligned}
    \hat{J}_{app, m} = & \int_0^{2\pi}\mathrm{d}\theta\ \int_{-\pi}^{\pi} \mathrm{d}\Delta \int_{\varphi=\pi/2}^{3\pi/2} \mathrm{d}\varphi\ \frac{1}{2\pi^2}\ |\sin(\Delta/2)| \cos\varphi \\* & \sum_{n_1, n_2}\hat{p}_{n_1}\hat{p}_{n_2} e^{\mathrm{i}(n_1+n_2-m)\theta}e^{\mathrm{i}n_2\Delta}.
\end{aligned}
\label{eq:derivation:app}
\end{equation}
By using $\beta=\theta+\Delta$, it is a straightforward integration that
\begin{equation}
    \hat{J}_{app, m} = -\frac{4}{2\pi}\sum_n \hat{p}_{m-n}\hat{p}_{n}
        \frac{1}{1/4-n^2}
\end{equation}

\subsection{Receding part}
The receding part of the integral reads in Fourier space
\begin{equation}
    \begin{aligned}
        &\hat{J}_{rec, m} = \int_0^{2\pi}\frac{\mathrm{d}\theta}{4\pi^2}e^{-\mathrm{i} m \theta} \int_{-\pi}^{\pi}\mathrm{d}\Delta\ \int_{-\pi/2}^{\pi/2} \mathrm{d}\varphi\ v_\mathrm{rel} \cos\varphi \\* 
        &\sum_{n_1, n_2}\hat{p}_{n_1}\hat{p}_{n_2} e^{\mathrm{i} n_1 \theta(t_0)}e^{\mathrm{i} n_2 \beta(t_0)} e^{-2 Sc J_s}
    \end{aligned}
    \label{eq:derivation:receding}
    \end{equation}
The integral $J_s$ is given by
\begin{equation}
\begin{aligned}
    J_s=\int_{t_0}^t \cos(\beta(t')-\theta(t')dt')=\frac{1}{2Sc}\ln\left(\frac{\mu}{\mu^2+1}\frac{1+\mu^2\exp(-4Sc t_{dur})}{\mu\exp(-2Sc t_{dur})}\right)
    \label{eq:derivation:js}
\end{aligned}
\end{equation}
where we again used the time dependence given by Eq.~\eqref{eq:derivation:angle} to calculate the integral.
The angles $\theta(t_0)$ and $\beta(t_0)$ can be calculated using Eq.~\eqref{eq:derivation:angle} and $\bar{c}=\text{const}$.
For now, we proceed by expanding these complicated expressions for small $Sc$.
First, substituting Eq.~\eqref{eq:derivation:tdur} into Eq.~\eqref{eq:derivation:js} gives
\begin{equation}
\begin{aligned}
    e^{-2Sc J_s} \approx 1 - 2 Sc \frac{\cos(\Delta)}{|\sin(\Delta/2)|} \cos\varphi + 2 Sc^2 \cos^2\varphi \frac{\mu^2-5}{1+\mu^2} + O(Sc^3).
\end{aligned}
\end{equation}
Similarly, substituting Eq.~\eqref{eq:derivation:tdur} into Eq.~\eqref{eq:derivation:angle} and expanding gives us
\begin{equation}
\begin{aligned}
    \Delta(t_0) \approx \Delta(t) + \sin(\Delta)\left[-Sc \frac{\cos\varphi}{|\sin(\Delta/2)|}-Sc^2\cos^2\varphi\right]+O(Sc^3) = \Delta(t)+A
\end{aligned}
\end{equation}
from which we find $\beta(t_0)\approx \beta(t)+A/2+O(Sc^3)$ and $\theta(t_0)\approx \theta(t)-A/2+O(Sc^3)$,
and therefore
\begin{equation}    
\begin{multlined}
        e^{\mathrm{i} n_1 \theta(t_0)}=e^{\mathrm{i} n_1 \theta(t)}\left(1+in_1\cos\varphi \frac{\sin\Delta}{|\sin(\Delta/2)|}Sc\right.\\* \left.+\left[in_1\sin\Delta\cos^2\varphi-\frac{n_1^2}{2}\frac{\sin^2\Delta}{\sin^2(\Delta/2)}\cos^2\varphi\right]Sc^2\right)
        \label{eq:derivation:exp1}
    \end{multlined}
    \end{equation}
    \begin{equation}
    \begin{multlined}
    e^{\mathrm{i} n_2 \beta(t_0)}=e^{\mathrm{i} n_1 \beta(t)}\left(1-in_2\cos\varphi \frac{\sin\Delta}{|\sin(\Delta/2)|}Sc\right.\\* \left.-\left[in_2\sin\Delta\cos^2\varphi+\frac{n_2^2}{2}\frac{\sin^2\Delta}{\sin^2(\Delta/2)}\cos^2\varphi\right]Sc^2\right)
\label{eq:derivation:exp2}
\end{multlined}
\end{equation}
Thus, the integrand of the receding integral Eq.~\eqref{eq:derivation:receding} becomes (using Eq.~\eqref{eq:derivation:vrel} for $v_\mathrm{rel}$), sorted by orders of $Sc$
\begin{equation}
\begin{aligned}
    \int_0^{2\pi}\mathrm{d}\theta \int_{-\pi}^{\pi}\mathrm{d}\Delta\ \int_{-\pi/2}^{\pi/2} \mathrm{d}\varphi\ \frac{1}{2\pi^2}\sum_{n_1, n_2}\hat{p}_{n_1}\hat{p}_{n_2}e^{\mathrm{i}(n_1+n_2-m)\theta}e^{\mathrm{i}n_2\Delta}\\*
    \bigg\{|\sin(\Delta/2)|\cos\varphi-Sc\cos^2\varphi\left[i(n_1-n_2)\sin\Delta+2\cos\Delta\right]\\*
    \left. Sc^2\cos^3\varphi\left[\left(-\frac{(n_1-n_2)^2}{2}-2\right)\frac{\sin^2\Delta}{|\sin(\Delta/2)|}+2i(n_2-n_1)\frac{\sin\Delta\cos\Delta}{|\sin(\Delta/2)|}\right. \right. \\* 
    \left.-2|\sin(\Delta/2)|\cos\Delta+i(n_1-n_2)|\sin(\Delta/2)|\sin\Delta{\bigg]}\right\}+O(Sc^3)
    \label{eq:derivation:recedingint}
\end{aligned}
\end{equation}
It is easily seen that the 0th order term cancels with the approaching part of the collision integral. After solving the remaining integrals, one obtains
\begin{equation}
    \begin{aligned}
        \hat{J}_{rec, m}+\hat{J}_{app, m} &= \sum_n \hat{p}_{m-n}\hat{p}_{n}\tilde{K}_{mn}
    \end{aligned}
    \label{eq:derivation:resultnoforbidden}
    \end{equation}
with the coupling matrix
\begin{equation}
\begin{aligned}
    \tilde{K}_{mn} = \frac{1}{2\pi}\left[-Sc\,m\,\pi^2\left(\delta_{n,1}-\delta_{n,-1}\right)+Sc^2\frac{8}{3}\left\{m\left(\frac{3/2m-n}{n^2-9/4}+\frac{1/2m+n}{n^2-1/4}\right)-8\right\}\right]
\end{aligned}
\end{equation}

\subsection{Contribution of the forbidden zone}
The integrand for the forbidden part is the same as in the receding part Eq.~\eqref{eq:derivation:recedingint}. All terms in Eq.~\eqref{eq:derivation:recedingint} $\sim Sc^2$ as well as the term $\sim Sc \sin(\Delta)$ do not contribute because they are at least of order $O(Sc^3)$ when integrated from $-\Delta_c$ to $-\Delta_c$. Therefore, this part of the integral reads
\begin{equation}
    \begin{aligned}
        &\hat{J}_{for, m} = \int_0^{2\pi}\mathrm{d}\theta\int_{-\pi/2}^{\pi/2} \mathrm{d}\varphi\ \int_{-\Delta_c}^{\Delta_c} \mathrm{d}\Delta\ \frac{1}{2\pi^2} \sum_{n_1, n_2}\hat{p}_{n_1}\hat{p}_{n_2}e^{\mathrm{i}(n_1+n_2-m)\theta}e^{\mathrm{i}n_2\Delta}\\*
        &\left(|\sin(\Delta/2)|\cos\varphi-2Sc\cos^2\varphi)\right) +O(Sc^3)
    \end{aligned}
    \end{equation}
As $\Delta$ is small in the forbidden zone and we are only interested in the correction of order $O(Sc^2)$, we can expand all functions around $\Delta=0$
\begin{equation}
\begin{multlined}
        \hat{J}_{for, m} = \int_0^{2\pi}\mathrm{d}\theta\int_{-\pi/2}^{\pi/2} \mathrm{d}\varphi\ \int_{-4Sc\cos\varphi}^{4Sc\cos\varphi} \mathrm{d}\Delta\ \frac{1}{2\pi^2} \sum_{n_1, n_2}\hat{p}_{n_1}\hat{p}_{n_2}e^{\mathrm{i}(n_1+n_2-m)\theta}e^{\mathrm{i}n_2\Delta}\\*
        \left(|\Delta/2|\cos\varphi-2Sc\cos^2\varphi)\right) +O(Sc^3)\\*
        =\frac{1}{2\pi}\frac{64}{3}Sc^2\sum_{n}\hat{p}_{m-n}\hat{p}_{n} +O(Sc^3)
        \label{eq:derivation:forbiddenresult}
\end{multlined}
\end{equation}

By adding Eq.~\eqref{eq:derivation:forbiddenresult} to Eq.~\eqref{eq:derivation:resultnoforbidden}, we arrive at our final result
\begin{equation}
    \partial_t \hat{p}_m - \frac{1}{2}(\nabla^* \hat{p}_{m-1}+\nabla \hat{p}_{m+1})= \rho\hat{J}_{\mathrm{coll}, m}
\end{equation}
\begin{equation}
    \begin{aligned}
        \hat{J}_{\mathrm{coll}, m} = \sum_n \hat{p}_{m-n}\hat{p}_{n}K_{mn}
    \end{aligned}
    \label{eq:derivation:finalresult}
    \end{equation}
with the coupling matrix
\begin{equation}
\begin{aligned}
    K_{mn} = \frac{1}{2\pi}\left[-Scm\pi^2\left(\delta_{n,1}-\delta_{n,-1}\right)+Sc^2\frac{8}{3}m\left(\frac{3/2m-n}{n^2-9/4}+\frac{1/2m+n}{n^2-1/4}\right)\right]
\end{aligned}
\end{equation}

\section{Solving the collision integral for non-reciprocal interactions}
\subsection{Two particle dynamics}
Analogous to the reciprocal case
\begin{equation}
    \dot{\Delta} = \dot{\beta}-\dot{\theta} = -Sc^+\sin(\theta-\beta)= Sc^+ \sin(\Delta),
\end{equation}
meaning that the difference in directions $\Delta$ still follows the same equations, with the replacement $2Sc \rightarrow Sc^+$.
We dropped the arguments of $Sc^\pm(\sigma,\tilde{\sigma})$ for notational simplicity.
The sum of the directional angles $s = \theta+\beta$ is conserved during a reciprocal interaction, but not for non-reciprocal interactions.
We define $\bar{c}$ as it's value at the time $t$ the collision ends.
\begin{equation}
    \bar{c} = s(t)
\end{equation}The following identities can be shown (see Appendix \ref{app:identnonrec})
\begin{eqnarray}
    v_\mathrm{rel} =&2 v_{0} \left| \sin \frac{\Delta}{2} \right|\\
    \label{eq:nonrec:vrel}
    \sin(c/2) =& -\cos(\varphi)\text{sgn}(\sin(\Delta/2))\\
    \label{eq:nonrec:sinhalfc}
    \cos(c/2) =& -\sin(\varphi)\text{sgn}(\sin(\Delta/2)),\\
    \label{eq:nonrec:coshalfc}
\end{eqnarray}
where $c=\bar{c}-2\varphi$.
The equation of motion for the sum of the angles is
\begin{equation}
    \dot{s}=Sc^-\sin\Delta
\end{equation}
and therefore
\begin{equation}
    s(\bar{t})=\bar{c}-Sc^-\int_t^{\bar{t}}\sin\Delta(t')\ dt'= \bar{c}-\frac{2Sc^-}{Sc^+}\left[\arctan(\mu\exp(-Sc^+(t-\bar{t})))-\arctan(\mu)\right].
    \label{eq:nonrec:sumappendix}
\end{equation}
See Appendix \ref{app:identnonrec} for the calculation of the integral.
And for the dynamics of the position, the following equation holds
\begin{equation}
\begin{aligned}
    \dot{\vec{\Delta r}} = \begin{pmatrix}
        \cos\beta-\cos\theta\\* \sin\beta-\sin\theta
    \end{pmatrix} = \begin{pmatrix}
        -2\sin(s/2)\sin(\Delta/2)\\* 2\cos(s/2)\sin(\Delta/2)
    \end{pmatrix}.
\end{aligned}
\end{equation}
The change in $s$ over the course of a collision is of order $O(Sc)$. We therefore expand to first order in $(s(\bar{t})-\bar{c})$.
\begin{equation}
    \begin{aligned}
        \dot{\vec{\Delta r}} = 2\left[\begin{pmatrix}
            -\sin(\bar{c}/2)\\* \cos(\bar{c}/2)
        \end{pmatrix}-\frac{1}{2}\begin{pmatrix}
            \cos(\bar{c}/2)\\* \sin(\bar{c}/2)
        \end{pmatrix}(s(\bar{t})-\bar{c})\right]\sin(\Delta/2) + 0(Sc^2).
    \end{aligned}
    \end{equation}
Integrating leads to 
\begin{equation}
\begin{aligned}
    \vec{\Delta r}(t_0)-\vec{\Delta r}(t) = \begin{pmatrix}
        -2\sin(\bar{c}/2)\\* 2\cos(\bar{c}/2)
    \end{pmatrix}G-\begin{pmatrix}
        \cos(\bar{c}/2)\\* \sin(\bar{c}/2)
    \end{pmatrix}H.
    \label{eq:nonrec:position}
\end{aligned}
\end{equation}
With $G$ and $H$ given by
\begin{eqnarray}
        &G=\int_t^{t_0}\sin(\Delta(t')/2)dt'=\frac{1}{-2Sc}\left(-\arcsinh(\mu\exp(-2Sc(t-t_0))+\arcsinh\mu)\right).\\
    \label{eq:derivation:gint}
   &H=\int_t^{t_0}\sin(\Delta(t')/2)\cdot(s(t')-\bar{c})dt'=\frac{2Sc^-}{(Sc^+)^2}\int_{\mu}^{\mu\exp(-Sc^+(t-t_0))}\frac{\arctan(x)-\arctan(\mu)}{\sqrt{1+x^2}}dx
    \end{eqnarray}
and $\mu=\tan(\Delta(t)/2)$. Here we substituted Eq.~\eqref{eq:derivation:angle} to calculate the integral.
Completely analogous to the reciprocal case, the next step is to bring $\vec{\Delta r}(t)$ to the other side of Eq.~\eqref{eq:nonrec:position} and square the equation. We look at collisions where $\vec{\Delta r}(t)=\begin{pmatrix}
    \cos\varphi& \sin\varphi
\end{pmatrix}^T$ and the collision ends at $t_0$ with $\vec{\Delta r}(t_0)^2=1$. The equation we need to solve to obtain the duration of the collision is therefore
\begin{equation}
\begin{aligned}
    0=4G^2+H^2-4\sin(c/2)G-2\cos(c/2)H.
    \label{eq:nonrec:criterium}
\end{aligned}
\end{equation}
We try the ansatz that $t_{dur}=t-t_0$ is of similar form as for reciprocal interactions, with a small correction $\delta$
\begin{equation}
    t_{dur}= -\frac{1}{Sc^+}\ln\left\{\frac{1}{\mu}\sinh\left[Sc^+\sin(c/2)+\arcsinh\mu+\delta\right]\right\}.
    \label{eq:nonrec:tdurappendix}
\end{equation}
Entering this into Eq.~\eqref{eq:nonrec:criterium} and expanding H for small $\delta$ gives (see Appendix \ref{app:identnonrec} for more details)
\begin{equation}
    4\delta+\frac{(Sc^-)^2}{(Sc^+)^2}\cos^2\frac{\Delta}{2}\left(Sc^+\sin\frac{c}{2}+\delta\right)^3+2Sc^-\cos\frac{\Delta}{2}\cos\frac{c}{2}\left(Sc^+\sin\frac{c}{2}+\delta\right)=0.
\end{equation}
Therefore we get approximately (already inserting Eq.~\eqref{eq:nonrec:sinhalfc} and Eq.~\eqref{eq:nonrec:coshalfc})
\begin{equation}
    \delta=-\frac{1}{2}Sc^-Sc^+\cos\frac{\Delta}{2}\sin\varphi\cos\varphi.
\end{equation}

The expression Eq.~\eqref{eq:nonrec:tdur} diverges (remembering identity Eq.~\eqref{eq:nonrec:sinhalfc} and Eq.~\eqref{eq:nonrec:coshalfc}) if
\begin{equation}
    -Sc^+\cos\varphi-\frac{1}{2}Sc^-Sc^+\cos\frac{\Delta}{2}\sin\varphi\cos\varphi+\arcsinh(\tan(|\Delta/2|))<0.
    \label{eq:nonrec:forbidden_condition}
\end{equation}
As the forbidden zone only contributes in second order in $Sc$ (we know this is true for the reciprocal case, and can later verify it for the non-reciprocal corrections), we can ignore the higher order correction to the forbidden zone and treat it like in the case with reciprocal interactions, with the replacement $2Sc\rightarrow Sc^+$.
Furthermore, we notice that the correction term $\delta$ will not contribute to the final result at all. Terms $\sim \delta$ are uneven in $\varphi$ (all other contributions only contain $\cos \varphi$ and are therefore even in $\varphi$) and will therefore vanish under the integral. Terms $\sim \delta^2$ are of higher than second order.
We will therefore go back to the expression without the $\delta$ in the following calculations.

\subsection{Evaluation}

In summary, the duration $t_{dur}$, the boundaries of the forbidden zone and the dynamics of $\Delta$ all stay the same as in the previous section (up to $O(Sc^2)$), with the replacement $2Sc\rightarrow Sc^+$. The only thing that changes is that $s$ is no longer constant.
Inserting Eq.~\eqref{eq:nonrec:tdur} into Eq.~\eqref{eq:nonrec:sum} and expanding gives
\begin{equation}
    \begin{aligned}
        s(t_0) \approx \bar{c} - 2\cos(\Delta/2)\left(Sc^-\sin(c/2)\right)+\cos(\Delta/2)\sin(\Delta/2)Sc^-Sc^+\sin^2(c/2)=\bar{c}+B
    \end{aligned}
    \label{eq:schange}
    \end{equation}
Analogous to the reciprocal case, one can calculate the pre-collisional angles from $\beta(t)\approx \beta(t)+A/2+B/2+O(Sc^3)$ and $\theta(t)\approx \theta(t)-A/2+B/2+O(Sc^3)$
and therefore
\begin{eqnarray}
        e^{\mathrm{i} n_1 \theta(t_0)}&=e^{\mathrm{i} n_1 (\theta(t)-A/2)}e^{B/2}\\
    e^{\mathrm{i} n_2 \beta(t_0)}&=e^{\mathrm{i} n_2 (\theta(t)+A/2)}e^{B/2}.
\end{eqnarray}
We end up with all terms from the reciprocal case (with the replacement $2Sc\rightarrow Sc^+$), multiplied with the following term
\begin{equation}
\begin{multlined}
    e^{\mathrm{i} (n_1+n_2) B/2}=1+i(n_1+n_2)\left(-\cos(\Delta/2)\left(Sc^-\sin(c/2)\right)\right.\\*
    \left.+\frac{1}{2}\cos(\Delta/2)\sin(\Delta/2)Sc^-Sc^+\sin^2(c/2)\right)
    -(n_1+n_2)^2\frac{1}{2}\cos^2(\Delta/2)\left(Sc^-\sin(c/2)\right)^2.
\end{multlined}
\end{equation}
Multiplying everything out, one obtains all terms from the reciprocal case and corrections for the non-reciprocal interactions
\begin{multline}
        \int_0^{2\pi}\mathrm{d}\theta \int_{-\pi}^{\pi}\mathrm{d}\Delta\ \int_{-\pi/2}^{\pi/2} \mathrm{d}\varphi\ \frac{1}{2\pi^2}\sum_{n_1, n_2}\hat{p}_{n_1}(\sigma)\hat{p}_{n_2}(\tilde{\sigma})e^{\mathrm{i}(n_1+n_2-m)\theta}e^{\mathrm{i}n_2\Delta}\\*
        \left\{+i(n_1+n_2)\sin(\Delta/2)\cos(\Delta/2)\cos^2\varphi (-Sc^-)\right.\\*
        \left.+i(n_1+n_2)\text{sgn}(\sin(\Delta/2))\cos(\Delta/2)\left[i(n_1-n_2)\sin\Delta-2\cos\Delta\right]\cos^3\varphi Sc^-Sc^+\right.\\*
        \left.+i(n_1+n_2)\cos(\Delta/2)\sin(\Delta/2)|\sin(\Delta/2)|\frac{\cos^3\varphi}{2} Sc^-Sc^+\right.\\*
        \left.-(n_1+n_2)^2\cos^2(\Delta/2)|\sin(\Delta/2)|\frac{\cos^3\varphi}{2}(Sc^-)^2\right\}
    \end{multline}
We note that these terms are not affected by the forbidden zone in first and second order in $Sc$.
After carrying out the integration, the resulting corrections to $\hat{J}^{\sigma\tilde{\sigma}}_{\mathrm{coll},m}$ are
    \begin{multline}
            \frac{2}{2\pi}\sum_{n_2}\hat{p}_{m-n_2}(\sigma)\hat{p}_{n_2}(\tilde{\sigma})
            \left\{\frac{\pi^2}{4}m(\delta_{n_2,1}-\delta_{n_2,-1})Sc^-\right.\\*
            \left.-\frac{16m(3m+3n_2-4mn_2^2)}{9-40n_2^2+16n_2^4}\frac{2}{3}Sc^-Sc^+\right.
            \left.-\frac{4m^2(3-4n_2^2)}{9-40n_2^2+16n_2^4}\frac{2}{3}(Sc^-)^2\right\}.
        \end{multline}

\subsection{Change of $S$ during a collision}
\label{app:add}
Equation \eqref{eq:schange} is equally valid if one exchanges the roles of $t$ and $t_0$, meaning $t$ is now the time at the start of the collision.
In this case $\varphi \in \left[-\pi/2,\pi,2\right]$, as these are the values possible at the start of a collision.
By keeping only the first order term of Eq. \eqref{eq:schange} and squaring the equation, one arrives at
\begin{equation}
    \delta s^2(\Delta,\varphi)=4\cos^2(\Delta/2)\cos^2(\varphi)(Sc^-)^2.
\end{equation}

\section{Identities and Integrals}
Most of the these evaluations were originally presented in Ref. \cite{ihle2023asymptotically} and are shown here again for the sake
of selfcontainedness.
\label{app:identnonrec}
\section*{Relative velocity}
The identity for $v_\mathrm{rel}$ is shown as follows:
\begin{equation}
    \begin{aligned}
    v_\mathrm{rel} &=v_{0}\left(\hat{n}_{2}-\tilde{n}_{1}\right) \\*
    v_\mathrm{rel}^{2} &=v_{0}^{2}\left[\left(\cos \theta_{2}-\cos \theta\right)^{2}+\left(\sin \theta_{2}-\sin \theta\right)^{2}\right] \\*
    &=v_{0}^{2}\left[-2 \cos \theta_{2} \cos \theta-2 \sin \theta_{2} \sin \theta+2\right] \\*
    &=v_{0}^{2}\left[-2 \cos \left(\theta_{2}-\theta\right)+2\right] \\*
    &=v_{0}^{2}\left[-2\left(1-\sin ^{2}\left(\frac{\Delta}{2}\right)\right)+2\right] \\*
    &=2 v_{0}^{2} \sin ^{2} \frac{\Delta}{2} \\*
    \rightarrow v_\mathrm{rel} &=2 v_{0} \left| \sin \frac{\Delta}{2} \right|.
    \end{aligned}
    \end{equation}
\section*{Identities for $\sin{c/2}$ and $\cos{c/2}$}
We have
\begin{equation}
    \begin{aligned}
    v_\mathrm{rel} \cos \varphi&=\vec{v}_\mathrm{rel}\hat{\Delta r}\\*
    &=v_{0}\left(\hat{n}_{2}-\hat{n}\right) \hat{\Delta r} \\*
    &=v_{0}\left(\cos (\beta-\varphi)-\cos (\theta_1-\varphi)\right)\\*
    &=v_{0}\left(\cos (\frac{c}{2}+\frac{\Delta}{2})-\cos (\frac{c}{2}-\frac{\Delta}{2})\right)\\*
    &=-2v_{0} \sin \frac{c}{2} \sin \frac{\Delta}{2}
\end{aligned}
\end{equation}
and therefore, using the identity for $v_\mathrm{rel}$,
\begin{equation}
    \begin{aligned}
    \Rightarrow&\sin(c/2) = -\cos(\varphi)\text{sgn}(\sin(\Delta/2))
    \end{aligned}
    \end{equation}
Here we used 
\begin{equation}
    \alpha_{i} := \theta_i-\varphi
\end{equation}
\begin{equation}
    \alpha_{1}=\frac{1}{2}\left[\alpha_{1}+\alpha_{2}+\left(\alpha_{1}-\alpha_{2}\right)\right]=\frac{c}{2}-\frac{\Delta}{2},\ \ \ \alpha_{2}=\frac{c}{2}+\frac{\Delta}{2}
\end{equation}
in the fourth equality.
We proceed analogously for the Cosine:
\begin{equation}
    \begin{aligned}
    v_\mathrm{rel} \sin \varphi&=\vec{v}_\mathrm{rel} \times \hat{\Delta r}\\*
    &=v_{0}\left(\hat{n}_{2}-\hat{n}\right) \times \hat{\Delta r} \\*
    &=v_{0}\left(\sin (\beta-\varphi)-\sin (\theta_1-\varphi)\right)\\*
    &=v_{0}\left(\sin (\frac{c}{2}+\frac{\Delta}{2})-\sin (\frac{c}{2}-\frac{\Delta}{2})\right)\\*
    &=-2v_{0} \cos \frac{c}{2} \sin \frac{\Delta}{2}
\end{aligned}
\end{equation}
and therefore
\begin{equation}
    \begin{aligned}
    \Rightarrow&\cos(c/2) = -\sin(\varphi)\text{sgn}(\sin(\Delta/2)).
    \end{aligned}
    \end{equation}

\section*{Sum of the angles}
To calculate Eq.~\eqref{eq:nonrec:sum}, we use the identity $\sin \Delta = 2\tan (\Delta/2)/(1+\tan^2(\Delta/2))$
\begin{equation}
    \begin{aligned}
    s(\bar{t})&=\bar{c}-Sc^-\int_t^{\bar{t}}2\tan (\Delta/2)/(1+\tan^2(\Delta/2))\ dt'\\*
    &=\bar{c}-Sc^-\int_t^{\bar{t}}2\mu\exp(Sc^+(t-t'))/(1+\mu^2\exp^2(Sc^+(t-t')))\ dt'
    \end{aligned}
\end{equation}
and substitute $x=\mu\exp(Sc^+(t-t'))$
\begin{equation}
    \begin{aligned}
    &=\bar{c}-Sc^-\int_{\mu}^{\mu\exp(Sc^+(t-\bar{t}))}-\frac{2}{Sc^+}\frac{1}{1+x^2}\ dx\\*
    &= \bar{c}-\frac{2Sc^-}{Sc^+}\left[\arctan(\mu\exp(Sc^+(t-\bar{t})))-\arctan(\mu)\right]
    \end{aligned}
\end{equation}
\section*{Integral H}
We insert all definitions and make the substitution $x=\mu\exp(Sc^+(t-t'))$
\begin{equation}
    \begin{aligned}
        H&=\int_t^{t_0}\sin(\Delta(t')/2)\cdot(s(\bar{t})-\bar{c})dt'\\*
        &=-\int_t^{t_0}\frac{\tan(\Delta(t')/2)}{\tan^2(\Delta(t')/2)+1}\cdot\frac{2Sc^-}{Sc^+}\left[\arctan(\mu\exp(Sc^+(t-t')))-\arctan(\mu)\right]dt'\\*
        &=\frac{2Sc^-}{(Sc^+)^2}\int_{\mu}^{\mu\exp(Sc^+(t-t_0))}\frac{\arctan(x)-\arctan(\mu)}{\sqrt{1+x^2}}dx
    \end{aligned}
\end{equation}
By inserting the ansatz for $t_{dur}$, we have
\begin{equation}
    \begin{aligned}
        H&=\frac{2Sc^-}{(Sc^+)^2}\int_{\mu}^{\sinh\left[-Sc^+\sin(c/2)+\arcsinh\mu+\delta\right]}\frac{\arctan(x)-\arctan(\mu)}{\sqrt{1+x^2}}dx
    \end{aligned}
\end{equation}
we substitute $x=\sinh y$ and get
\begin{equation}
    \begin{aligned}
        H&=\frac{2Sc^-}{(Sc^+)^2}\int_{\arcsinh\mu}^{\arcsinh\mu-Sc^+\sin(c/2)+\delta}(\arctan(\sinh(y))-\arctan(\mu)) dy
    \end{aligned}
\end{equation}
which can easily be expanded for small $-Sc^+\sin(c/2)+\delta$.
\begin{equation}
    \begin{aligned}
        H&\approx\frac{2Sc^-}{(Sc^+)^2}\left(-Sc^+\sin(c/2)+\delta\right)^2\cos(\Delta/2)
    \end{aligned}
\end{equation}

\end{document}